\documentclass[a4paper]{article}

\usepackage{mystyle}

\usepackage[utf8]{inputenc} 
\usepackage[T1]{fontenc}    
\usepackage{hyperref} 
\usepackage{gensymb}
\usepackage{indentfirst}
\usepackage{graphicx}
\usepackage{epstopdf}
\usepackage{url}            
\usepackage{booktabs}       
\usepackage{amsfonts}       
\usepackage{nicefrac}       
\usepackage{microtype}      
\usepackage{lipsum}
\usepackage{xcolor}
\usepackage{subcaption}
\usepackage{environ}
\usepackage{pifont} 
\usepackage{enumitem}
\usepackage{doi}
\usepackage{comment}
\usepackage{titlesec}
\usepackage{amsmath, amsfonts, amssymb}
\usepackage{tabularx,ragged2e,booktabs,caption} 
\newcolumntype{C}[1]{>{\Centering}m{#1}}

\usepackage{titlesec}
\usepackage{mdframed}
\usepackage{caption}
\usepackage{placeins}    

\usepackage[backend=biber,style=numeric,sorting=none,maxnames=3,minnames=1]{biblatex}
\usepackage{listings}
\lstdefinestyle{fortranstyle}{
    basicstyle=\ttfamily\footnotesize,
    backgroundcolor=\color{bg},  
    frame=single,        
    rulecolor=\color{black}, 
    keywordstyle=\color{blue},
    commentstyle=\color{gray},  
    stringstyle=\color{red},
    showstringspaces=false,  
    breaklines=true,    
    escapeinside={@}{@}, 
    numbers=none,   
    language=Fortran, 
    columns=fixed,  
    basewidth=0.5em                  
}
\definecolor{bg}{rgb}{0.95,0.95,0.95} 
\definecolor{lightmagenta}{rgb}{1.0, 0.8, 1.0}

\definecolor{myYellow}{rgb}{1, 1, 0.1}
\definecolor{myBlue}{rgb}{0.4, 0.55, 1}
\newmdenv[
  backgroundcolor=myYellow,
  skipabove=10pt,
  skipbelow=10pt,
  leftmargin=0pt,
  rightmargin=0pt,
  innertopmargin=5pt,
  innerbottommargin=5pt,
  innerleftmargin=5pt,
  innerrightmargin=5pt,
  roundcorner=5pt
]{notResolved}
\newmdenv[
  backgroundcolor=myBlue,
  skipabove=10pt,
  skipbelow=10pt,
  leftmargin=0pt,
  rightmargin=0pt,
  innertopmargin=5pt,
  innerbottommargin=5pt,
  innerleftmargin=5pt,
  innerrightmargin=5pt,
  roundcorner=5pt
]{isResolved}

\newmdenv[
  backgroundcolor=lightmagenta,
  skipabove=10pt,
  skipbelow=10pt,
  leftmargin=0pt,
  rightmargin=0pt,
  innertopmargin=5pt,
  innerbottommargin=5pt,
  innerleftmargin=5pt,
  innerrightmargin=5pt,
  roundcorner=5pt
]{editbox}

\renewcommand{\thesection}{\arabic{section}.}
\renewcommand{\thesubsection}{\thesection\arabic{subsection}.}
\renewcommand{\thesubsubsection}{\thesubsection\arabic{subsubsection}.}

\titleformat{\section}[hang]
  {\normalfont\Large\bfseries}
  {\thesection}{0.5em}{}

\titleformat{\subsection}[hang]
  {\normalfont\large\itshape}
  {\thesubsection}{0.5em}{}

\titleformat{\subsubsection}[hang]
  {\normalfont\large\itshape}
  {\thesubsubsection}{0.5em}{}

\titlelabel{\thetitle.\quad}

\title{MARFA: An Effective Line-by-line Tool for Calculating Molecular Absorption in Planetary Atmospheres}

\author{
    \textbf{Mikhail Razumovskiy} \thanks{Corresponding author} \\
    Moscow Institute of Physics and Technology, Moscow\\
    \texttt{mrazumovskyy@gmail.com}
    \and
    \textbf{Boris Fomin}\\
    Central Aerological Observatory, Moscow\\
    \texttt{b.fomin@mail.ru}
    \and
    \textbf{Denis Astanin}\\
    Nuclear University MEPhI, Moscow\\
    \texttt{densof161922@gmail.com}
}

\renewcommand{\shorttitle}{\textit{MARFA: an effective tool for calculating molecular atmospheric absorption} }

\hypersetup{
pdftitle={A template for the arxiv style},
pdfsubject={astro-ph.EP},
pdfauthor={Mikhail Razumovskiy},
pdfkeywords={Venus, atmospheric absorption, PT-table, Fortran},
}

\titlespacing*{\subsection}
  {0pt}{*1}{*0.5}

\begin{document}
\maketitle
\large
\begin{abstract}
We present MARFA (Molecular atmospheric Absorption with Rapid and Flexible Analysis) -- an open-source line-by-line tool for calculating absorption coefficients and cross-sections in planetary atmospheres, particularly under conditions of uncertain spectroscopic data and missing continuum functions. With incorporated eleven-grid interpolation technique MARFA shows good performance in computation of far-wing contributions for large line cut-offs. The tool supports flexible parameterization, including line shape functions, wing corrections, user-defined atmospheric profiles, thus, facilitating rapid sensitivity studies for sparse datasets. Spectra are calculated at a high-resolution of about $5\cdot10^{-4}$ cm$^{-1}$, optimized for infrared and visible spectral regions where HITRAN-formatted line data is available, yet adaptable to other datasets with available line parameters. Output is represented either in a form of binary lookup tables files, directly compatible with radiative transfer codes or in a human-readable format for data analysis and distribution. The MARFA tool is provided in two ways: through a web application accessible at \href{https://marfa.app}{\textit{marfa.app}} for onboarding and educational usage, and as an open-source code available in a public \href{https://github.com/Razumovskyy/MARFA}{repository} for advanced utilization, development and contributions.
\end{abstract}


\section{Introduction}
\label{sec:intro}

\qquad Since 1960s \parencite{Drayson1966} a significant number of line-by-line Radiative Transfer Models (RTMs) have been developed to model radiative transfer in atmospheres of Earth and other terrestrial planets. Line-by-line approach implies that molecular absorption is treated very accurately by taking into account per-line data. Applications include remote sensing of planetary atmospheres, as well as reference radiative flux calculations necessary for weather and climate modeling. Precise atmospheric absorption calculations are essential in climate and weather studies and in satellite remote sensing of the atmosphere and surface.

\qquad One of the most notable models in this domain is the LBLRTM model \parencite{LBLRTM1992, LBLRTM2005}, which is central to the line-by-line methodology and has set standards for next generations of RTMs. In conventional RTMs like LBLRTM, absorption spectra are usually obtained as intermediate values before solving radiative transfer equations. Overall process roughly involves three stages: (1) line-by-line summation of absorption profiles at required atmospheric levels and spectral intervals, (2) incorporation of non-resonant continuum contributions (e.g. colision induced absorption) and optionally convolution with apparatus functions and (3) solving radiative transfer equations to derive radiances, transmittances, fluxes or heating rates. While this workflow is proven to be an effective strategy for Earth-like conditions, environments where continuum function is poorly characterized - as is the case e.g. for Venus \parencite{Haus2010, Haus2015} - require considering extended line cut-offs. This requirement dramatically increases the costs at stage (1). To target this bottleneck, we have developed MARFA code, which can be understood as a flexible framework for (re-)calculation of molecular atmospheric absorption in data-sparse or uncertain data environments.

\qquad MARFA (Molecular atmospheric Absorption with Rapid and Flexible Analysis) is a line-by-line tool optimized for sensitivity analysis of molecular atmospheric absorption, focused on rapid recalculation under uncertain spectroscopic and atmospheric conditions. It is a modern Fortran, open-source code with modular architecture that incorporates an eleven-grid interpolation technique \parencite{fomin1995effective} in its core. Application of this interpolation algorithm allows  handling large line cut-offs (up to 500 cm$^{-1}$ and beyond if needed) quickly and accurately. MARFA supports flexible parameterization of line shapes, sub-Lorentz corrections, user-defined atmospheric profiles and seamless switching between spectral databases. Its modular design and ability to introduce user-defined features allow users to utilize MARFA as the kernel in their own molecular absorption calculation setups on local computers, without relying on external tools that may not accommodate all use cases (e.g. extreme exoplanetary scenarios). Output values are calculated for all atmospheric levels at once and are provided in a binary pt-format, so afterwards they can directly serve as inputs to radiative transfer codes. The legacy version of MARFA has been tested in remote sensing applications \parencite{Fomin2020}, and its efficiency in line-by-line calculations has been independently verified \parencite{Kuntz1999}.

\qquad Particular attention has been given to present codes in a modern and accessible way on the GitHub \href{https://github.com/Razumovskyy/MARFA}{page}, tailored for the scientific and IT communities. The project is well documented in two ways: in-code and outside the code. Within the codebase, we provide extensive comments and clear variable names and their descriptions. Legacy code has very high refactoring coverage making it easier to navigate. Outside the code we provide supporting and comprehensive README file. It includes explanation of installation process and instructions for providing custom features. As with any code project, MARFA is not without areas for optimization and enhancement, but these are transparently listed on the GitHub \href{https://github.com/Razumovskyy/MARFA/issues}{issues page}, providing clarity for anyone interested in working with or contributing to the project. We have placed particular emphasis on organizing MARFA according to best practices of open-source community-driven projects.

\qquad As an addition to the MARFA source code which is the main result of this study, we present a web platform \href{https://marfa.app}{marfa.app} on which it is possible to interact with a light-weight version of MARFA. On this platform it is possible to perform molecular absorption calculations, generate plots and download data in both binary and human-readable formats. Additionally, it can serve as a valuable educational tool. Though not as rich in functionality as similar platforms (e.g. \parencite{HITRAN_on_the_WEB}), it is build on a modern, highly-scalable programming technologies stack exemplifying the possible direction of future scientific web-based labs. The code for the web platform is also publicly available and can be contributed \parencite{MARFA_webapp}.

\qquad The layout of this paper is as follows. In the next section, we explain the physical and computational basis under the core MARFA functionality. It includes brief overview of molecular absorption basis, interpolation technique and discussion of the implemented spectroscopic features. Section 3 provides a specific application scenario for the tool and focuses on the atmosphere of Venus. We show in this section why and how Venus atmosphere is taken as a benchmark. Section 4 discusses the MARFA code design and provides guidance on implementing custom features and outline the tool's limitation. In subsection 4.6, we provide details on the implementation of the web platform, followed by a few illustrative examples. In Section 5 we give some conclusive remarks and suggestions for future releases.

\section{Absorption spectra calculation framework}
\label{sec:methodology}

\qquad Considering specifics of calculation of atmospheric absorption in data-sparse environments, our framework prioritizes two main points. Firstly, we aim to equip users with an efficient line-by-line algorithm, minimizing concerns about spectral interval size and line cut-off condition and allowing fast recalculations of spectra on users' local computers. Secondly, we aim to provide an easily accessible tool by eliminating data hardcoding and to enable smooth parameter input and sensitivity analysis.

\subsection{Background}
\label{subsec: background}

\qquad Following the notation by \parencite{Goody1989}, the volume absorption coefficient in the atmosphere can be found by summing the contributions from individual lines across species ($l$~index) and transitions ($j$~index), and then add the continuum function: 
\begin{equation}
k_{\nu,\nu} = \sum_{j,l} n^l S_{n,j}^l f_j^l (\nu - \nu_{0,j}^l) + k_{\nu}^c.
\label{eq:goody_absorption}
\end{equation}
Our scheme is designed such that a separate calculation is required for each atmospheric species, and the continuum function estimation is not included. Resulting value might be either the volume absorption coefficient $k$ (typically in km$^{-1}$ for atmospheric modeling) or the absorption cross-section $\sigma$ in cm$^2$/molecule, so the expression~\ref{eq:goody_absorption} transforms into:
\begin{equation}
    k(\nu; p,T) = n(p,T) \sum_{j} S_j(T)f_j(\nu-\nu_0; \gamma(p,T))
    \label{eq:absorption_coefficient}
\end{equation}
\begin{equation}
    \sigma(\nu;p,T) = \sum_{j} S_j(T)f_j(\nu-\nu_0; \gamma(p,T))
    \label{eq:cross-section}
\end{equation}

In expressions~\ref{eq:absorption_coefficient} and~\ref{eq:cross-section} $S$ represents a temperature-dependent line intensity, while $f$ is a normalized line shape function, which depends on both temperature and pressure through the half-width at half-maximum (HWHM) $\gamma$. The parameter $\nu_0$ is a center of the line (wavenumber of a transition). The number density $n$ of the species can be determined, for example, from the atmospheric profile and the species' abundance (usually in ppm). Since the number density $n$ in expression~\ref{eq:absorption_coefficient} is outside the sum, the program can simply switch between calculation of absorption coefficient \ref{eq:absorption_coefficient} and cross-section \ref{eq:cross-section}. Initial per-line spectroscopic data, normally taken from spectroscopic databases like HITRAN \parencite{Gordon2022, Gordon2017}, HITEMP \parencite{HITEMP2010} or GEISA \parencite{GEISA2004, GEISA2008}, is required to determine entities under the summation sign in expressions~\ref{eq:absorption_coefficient} and~\ref{eq:cross-section}.

\subsection{Predefined line shapes}
\label{subsec: molecular absorption}
\qquad Within MARFA we have implemented the Lorentz, Gaussian and Voigt profiles which correspond to pressure-broadening, Doppler broadening and the combination of both effects respectively. Since the primary goal is a calculation of atmospheric absorption, Voigt line shape is set as a default choice. Details of our programmatic implementation of the Voigt function can be found below and in Appendix C.

\qquad Predefined Lorentz and Gaussian lines shapes are implemented exactly as per the well-known formulas:
\begin{equation}
    f_{\text{L}}(\nu;\, \gamma_L) = \frac{1}{\pi} \frac{\gamma_{\text{L}}}{\gamma_{\text{L}}^2 + (\nu - \tilde{\nu_{0}})^2} \qquad 
\end{equation}
\begin{equation}
f_{\text{G}}(\nu;\, \alpha_D) = \sqrt{\frac{\ln 2}{\pi \alpha_D^2}} \exp \left( - \frac{(\nu - \nu_{0})^2 \ln 2}{\alpha_D^2} \right), \qquad 
\end{equation}

where $\gamma_L=\gamma_L(P,T)$ and $\alpha_D=\alpha_D(T)$ are the corresponding line half-widths and $\tilde{\nu_0}$ is a pressure-shifted line center position.

\qquad Voigt line shape is described by the convolutional integral of Lorentzian and Gaussian line shapes:
\begin{equation}
    f_{\text{V}}(\nu - \nu_0, \gamma_L, \alpha_D) = f_{\text{L}} \otimes f_{\text{G}} = \int_{-\infty}^{\infty} \mathrm{d}\nu' \, f_{\text{L}}(\nu - \nu', \gamma_L) \times f_{\text{G}}(\nu' - \nu_0, \alpha_D)
\end{equation}

It can be effectively represented using the Voigt function $K(x,y)$ \parencite{Armstrong1967}:
\begin{equation}
f_{\text{V}}\left(\nu - \nu_0; \gamma_{\text{L}}, \alpha_{\text{D}}\right) = \frac{\sqrt{\ln 2 / \pi}}{\alpha_{\text{D}}} K(x, y),
\label{eq:voigt_line_shape}
\end{equation}
\begin{equation}
K(x, y) = \frac{y}{\pi} \int_{-\infty}^{\infty} \frac{e^{-t^2}}{(x - t)^2 + y^2} \, dt,
\label{eq:voigt_function}
\end{equation}
where $x = \frac{\sqrt{\ln 2} (\nu - \nu_0)}{\alpha_\text{D}}$ and $y = \frac{\sqrt{\ln 2} \, \gamma_\text{L}}{\alpha_\text{D}}$.

\qquad Many studies have proposed routines for determining the Voigt function through various numerical expansions, mainly due to its close relation to the complex error function. MARFA builds on the implementation of the \textcite{Humlicek1982} algorithm, optimized by \textcite{Kuntz1997} by reducing the number of floating-point operations by considering only the real part of the rational approximation. Additionally, since the primary use case for MARFA is extended line wings, we found it convenient to exploit the asymptotic convergence of Voigt to Lorentz for large $x$. We have ensured that the errors introduced by these asymptotic substitutions remain below 1\%. For more details, please refer to \hyperref[appx: Voigt]{Appendix C}. The legacy version of this approach has been successfully adopted in several RTMs and has been highlighted in radiation codes intercomparison research \parencite{Oreopoulos2012continual, Halthore2005intercomparison}. Our implementation of the Voigt function does not compete for peak precision, but it is efficient and perfectly suitable for setups with large line cut-offs. It should be noted that our focus is not on achieving the highest accuracy in out-of-the-box line shapes but rather on equipping users with the flexibility to introduce their own.

\qquad A user can manually introduce custom line shape function by creating a dedicated Fortran function (see documentation \parencite{MARFA_repo} for more details). In expressions~\ref{eq:absorption_coefficient} and~\ref{eq:cross-section}, the line shape function $f_j$ is always multiplied by the intensity $S_j$. To ensure compatibility, custom line shapes should follow this combined structure $S_j \times f_j$ inside their Fortran function implementations. For example, one needs to include the function at the end of the \texttt{Shapes.f90} module:

\begin{lstlisting}[style=fortranstyle]
real function myLineShape(X)  
    real(kind=DP), intent(in) :: X  ! Distance from line center [cm@\(^{-1}\)@] in double precision
    real :: myLineShape  
    ! ... User-defined line shape logic
    myLineShape = myLineShape * intensityOfT(temperature)  ! Mandatory @\(S_j(T)\)@ scaling  
end function myLineShape  
\end{lstlisting}

and redirect the module's shape function pointer:
\begin{lstlisting}[style=fortranstyle]
shapeFuncPtr => myLineShape  
\end{lstlisting}

\texttt{IntensityOfT}, implemented in the \texttt{Spectroscopy.f90} module, calculates the temperature-dependent line intensity using HITRAN's formulations \parencite{hitran_definitions_units}. Potentially, it can be also substituted with user's defined line intensity function. This design allows custom integration of advanced line profiles (e.g. \textcite{Tran2013, Ngo2013, Rosenkranz1985, Varghese1984}), without disrupting the main program workflow. Users can easily switch between different line shapes just by \texttt{shapeFuncPtr} pointer redirection.

\qquad The \texttt{Spectroscopy.f90} and \texttt{Shapes.f90} modules serve different roles. \texttt{Shapes.f90} contains available line shapes functions and governs default line shape selection. It must be noted that the choice of the line shape can significantly affect performance. The \texttt{Spectroscopy.f90} module contains standardized routines for calculation of temperature- and pressure- dependent parameters: Lorentz ($\gamma_{\text{L}}$) and Doppler ($\alpha_{\text{D}}$) half-widths, line intensities $S(T)$ and pressure-induced shifts\, $\tilde{\nu_0}$ \, following HITRAN's formulations \parencite{hitran_definitions_units}. Both of the modules are opened for extensibility: by users adding custom functionality and by future MARFA releases.

\subsection{$\chi$-factors}
\label{subsec: chis}

\qquad Experimental data suggest that the far wings of spectral lines deviate from Lorentz or Voigt behavior. To address this, usually so-called $\chi$-correction functions \parencite{Burch1969, Pollack1993} are introduced:
\begin{equation}
    f_{\text{corrected}}(\nu-\nu_0) = f_{\text{L}}(\nu-\nu_0)\chi(\nu-\nu_0),
\end{equation}

 The $\chi$-factor is an empirical function typically tied to a specific molecule, and the underlying physical mechanisms governing its absorption behavior. Applying such a correction improves physical reliability and accuracy of the results without compromising performance, because it doesn't require direct computations of underlying mechanisms (e.g. line mixing or finite duration of collisions). Depending on the molecule and specific atmospheric conditions, spectral line wings can exhibit either sub-Lorentzian behavior (e.g., CO$_2$ \parencite{Burch1969}) or super-Lorentzian behavior (e.g., H$_2$O \parencite{Clough1989}). For example, Tonkov correction \parencite{Tonkov1996} to the Lorentzian shape $f_{\text{L}}$ is implemented according to the piece-wise correction function: 
\begin{equation}
    \chi_{\text{Tonkov}} =
\begin{cases} 
    1.084 \exp(-0.027 \;|\nu-\nu_0|) & \text{if } 3\; \text{cm$^{-1}$} \leq |\nu-\nu_0| \leq 150\; \text{cm$^{-1}$}, \\
    0.208 \exp(-0.016 \;|\nu-\nu_0|) & \text{if } 150\; \text{cm$^{-1}$} < |\nu-\nu_0| \leq 300\; \text{cm$^{-1}$}, \\
    0.025 \exp(-0.009 \;|\nu-\nu_0|) & \text{if } |\nu-\nu_0| > 300\; \text{cm$^{-1}$}, \\
    1 & \text{otherwise}.
\end{cases}
\end{equation}

 \qquad Since we used Venus for benchmark studies (see Section~\ref{sec:Venus} for more details), we have implemented within MARFA several $\chi$-factors for CO$_2$ line shapes relevant to Venus atmospheric absorption modeling: \textcite{Tonkov1996}, \textcite{perrin1989chi-factor}, \textcite{Pollack1993}. Similar to the \texttt{Shapes.f90} module, \texttt{ChiFactors.f90} module contains implemented $\chi$-corrections. It is advisable to add custom-defined functions at the end of this module. Below is an example for users, how to add their own wings corrections functions by interacting with the MARFA codebase, \texttt{ChiFactors.f90} module:

\begin{lstlisting}[style=fortranstyle]
    pure function myChiFunction(X, moleculeIntCode) result(myChiFactor)
        implicit none
        real :: myChiFactor
        real(kind=DP), intent(in) :: X ! Distance from line center [cm@\(^{-1}\)@] in double precision
        integer, intent(in) :: moleculeIntCode ! e.g. 1 - H2O, 2 - CO2, ...

        myChiFactor = 1. ! default value
        
        ! set the molecule int code limitation 
        if (moleculeIntCode == 2) then 
            ! @$\chi$@-factor correction logic here 
        end if
    end function myChiFunction
\end{lstlisting}

By default, the \texttt{shapeFuncPtr} pointer directs to the \texttt{noneChi} function, meaning no correction applied. To change that, just redirect the module's $\chi$-function pointer:
\begin{lstlisting}[style=fortranstyle]
shapeFuncPtr => myChi
\end{lstlisting}
MARFA documentation \parencite{MARFA_repo} provides additional instructions on how to manually add custom $\chi$-corrections.

\subsection{Interpolation technique for the line-by-line modeling}
\label{subsec: line_summation}

\qquad Apart from improved algorithmic approaches for the Voigt function, interpolation techniques are essential for speeding-up line-by-line modeling. Notable works in this domain are: \textcite{Clough1989, Edwards1988, West1990, Gordley1994, Sparks1997, fomin1995effective, Titov1997, Kuntz1999, Kruglanski2005, Schreier2006}. One effective technique for addressing this is the so-called multi-grid approach. It makes estimations of a line profile function on a sequence of grids with increasing step sizes. The concept behind these algorithm is straightforward: moving away from the center of a spectral line results in the profile becoming progressively smoother, as in the case of a Lorentzian wing, where the profile follows $\sim 1/(\nu - \nu_0)^2$. The line profile is divided into sections corresponding to the number of grids. On the finest grid, the central, sharpest parts of the lines are summed. On the next, coarser grid, the smoother parts adjacent to the center are summed, and so on. Once all the parts of the lines have been summed on their respective grids, the discrete functions obtained on these grids are interpolated onto the finest grid, resulting in the desired spectrum.

\qquad In MARFA we use a multi-grid interpolation technique described by \textcite{fomin1995effective}, which is considered fast and suitable for calculations with various line contour models \parencite{Kuntz1999}. We have confirmed that in presence of large line cut-offs, 2 or 3 grids may not be enough to achieve the desired balance between speed and accuracy. To address this, our scheme employs 11 grids with increasing resolution. Interpolation technique is performed on 10 cm$^{-1}$-length intervals. Number of points in the finest grid is set to $N=20481$ leading to a resolution of $10/(N-1)\approx4.88\cdot10^{-4}\text{cm}^{-1}$. This is usually sufficient to resolve typical Doppler lines in upper atmosphere in IR-region. The number of points in other grids follow a rule: $N_i=10\cdot2^{i-1}+1$, $i=1, 10$ which by our experience is sufficient to describe distant line wings with line cut offs up to $500\;\text{cm}^{-1}$ and beyond. Important fact is that the interpolation requires significantly less computational time compared to calculation of line profiles on each grid. Interpolations between coarse and fine grids are performed only once (per subinterval $10 \;\text{cm}^{-1}$ step) and are independent of the number of lines, making the computation time for spectra logarithmically dependent on the line's half-width. The technique and its accuracy characteristics are detailed in \textcite{fomin1995effective} original study, which also discusses utilization of 3-point and 5-point interpolations in molecular spectra calculations.

\begin{figure}
  \centering
    \includegraphics[scale=0.45]{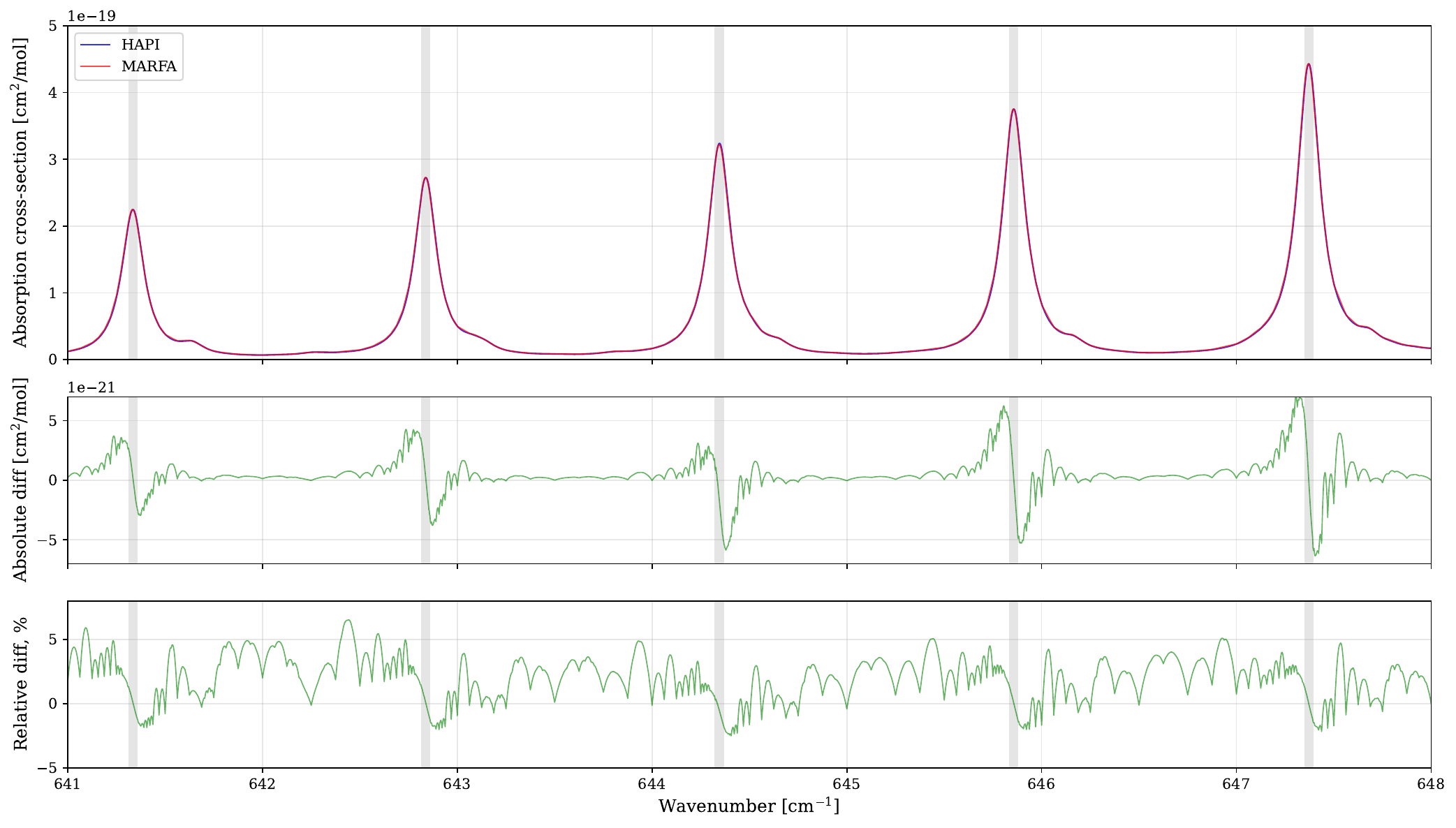}
  \caption{CO$_2$ absorption cross-sections calculated with HAPI and MARFA under conditions: $P$ = 1 atm, $T$ = 296 K, Lorentz line profile, 25 cm$^{-1}$ line cut-off and air as a broadening gas}
  \label{fig:hapi-marfa}
\end{figure}

\qquad To demonstrate speed and accuracy characteristics of MARFA, which in current version employs 3-point interpolations, we performed series of program runs to compare cross-sections computed with MARFA and with HAPI \parencite{HAPI} Python library. An example comparison for CO$_2$ is illustrated on fig. \ref{fig:hapi-marfa}. HAPI is known for its seamless integration with HITRAN data and proven reliability of results. When running calculations with HAPI we selected a default equidistant wavenumber grid, avoiding reliance on interpolation techniques, which inherently introduce numerical errors. Spectral resolution in both calculations was kept the same $~\sim5\cdot10^{-4}$ cm$^{-1}$. For the pressure of 1 atm this resolution is excessive but it gave us opportunity to estimate computational times for the most time-consuming spectra in the upper atmosphere. We found out that MARFA's computation is faster at around 2 orders of magnitude. In the line centers the discrepancies appear to be less than 1\%. However, the relative errors shown on fig. \ref{fig:hapi-marfa} in cross-section calculations between spectral lines can reach ~5\%. These errors can be reduced by an order of magnitude through the use of 5-point interpolation schemes instead of 3-point ones, with only a twofold decrease in computational speed \parencite{fomin1995effective}. But as seen from fig. \ref{fig:hapi-marfa}, the errors are oscillatory and, as practice shows, tend to cancel out each other to a large extent when integrated over the wavenumber. This case is relevant in the computation of integrated radiative fluxes or in the convolution of spectra with instrumental functions in remote sensing applications. These types of problems are that MARFA is primarily designed to address. Reference \parencite{fomin1995effective} further shows that the integration errors over the line profiles correspond to uncertainties in the line intensity parameter $S$ which at minimum is of ~1\%. For these reasons, the current version of MARFA keeps the 3-point interpolation scheme.

\qquad MARFA currently support a more limited set of spectroscopic options and line shapes compared to HAPI and does not offer direct integration of HITRAN data. However, it provides several advantages. These include: significantly higher calculation speed, support of calculation of absorption coefficients at all atmospheric levels at once, and enhanced functionality for accounting line wings contributions. Since primary aim of MARFA development was to target atmospheric studies for terrestrial planets, and not for purely spectroscopic purposes in Earth' environment, we view these two tools as complementary.

\section{Venus as a benchmark scenario}
\label{sec:Venus}

\subsection{Why Venus ?}
\qquad The idea of developing a separate tool for calculating molecular spectra originated during our ongoing work on constructing k-distribution parameterizations suitable for Venus GCMs \parencite{Fomin2022}, and particularly in the IR and visible spectral intervals (in progress). During this research, we faced considerable uncertainties in Venus' spectroscopic and atmospheric parameters to provide as inputs for forward radiative transfer model. That led to a need of rapidly recalculating absorption spectra for testing, validating and sensitivity analysis purposes. In Section~\ref{subsec:venus data limitations}, we provide a brief overview of the key uncertainties in the Venus' data essential for obtaining reliable outcomes from radiative transfer models.

\qquad In our view, Venus is a good starting point for validation and testing of new, general purpose line-by-line models, which aim to follow a data-driven and flexible approach. Despite decades of ground-based and spacecraft observations that yield substantial information on Venus, when it comes to radiative transfer modeling, reliable atmospheric and spectroscopic data to serve as inputs to forward models, remain limited. As RTMs are highly sensitive to atmospheric and spectroscopic parameters \parencite{Haus2015}, it is important that these models treat such data strictly as inputs, rather than hardcoding them. This is one of the major drawbacks why many Earth-based or data-specific models struggle to accurately capture Venusian conditions. 

\qquad Venus' unique rotation characteristics and its location within the habitable zone provide a clear analogy to a telluric exoplanet \parencite{Ehrenreich2012, Schaefer2011}. This makes it a promising test case for validating exoplanetary radiative transfer models. In exoplanetary research, radiative transfer models are highly sensitive to the input molecular absorption data. MARFA addresses this need by providing molecular spectra flexibly and precisely.

\subsection{Challenges and uncertainties in Venus atmosphere modeling}
\label{subsec:venus data limitations}

\qquad For each atmospheric constituent of interest, pre-calculated absorption coefficients are prepared for a set of pressure-temperature (P-T) value pairs. Significant differences in thermal structure with respect to latitude and local time (solar longitude) are observed in models of Venus's middle atmosphere \parencite[]{Zasova2006, Seiff1985}. Temperature profiles in the upper atmosphere differ drastically between the day- and night-side hemispheres \parencite[]{Keating1980upperVenus, keating1985models, seiff1982structure}. The temperature structure of the lower atmosphere is not well-studied and is often assumed to be homogeneous across the planet (globally averaged) in models. Due to this reason, interpolation and averaging procedures are employed and that leads to multiple approaches for compiling an a priori initial temperature profile for retrieval procedures \parencite{Tsang2009, Haus2013selfconsistent}. MARFA enables seamless switching between input atmospheric profiles (with up to 200 levels each), rather than just P-T values like conventional methods. It is useful when handling multiple atmospheric profile files and conducting comparative analyses. Radiative heating and cooling rates are usually calculated up to a height of 150 km \parencite{Haus2015}. This means that constituents mixing ratio profiles should also extend to this altitude. Most models use data compiled from various sources and if lacking data, interpolate to constant values in the upper atmosphere \parencite{Tsang2008, Pollack1993, Haus2015}.

\qquad The selected database for line parameters greatly affects the absorption cross-section and the resulting radiance spectra. Unfortunately, there is no dedicated Venus database with line parameters for all relevant species that accounts for CO$_2$
broadening and high-temperature conditions. As a result, line parameters for Venus modeling are typically taken from various databases and dedicated experimental studies \parencite{Haus2010, Haus2013selfconsistent, Haus2015}. Commonly used databases include HITRAN \parencite{RothmanHITRAN2004, rothman2008HITRAN2008, Rothman2013, Gordon2017, Gordon2022}, HITEMP \parencite{rothman1995hitranhitemp, HITEMP2010}, CDSD \parencite{tashkun2003cdsd}, and specific CO$_2$ data from \mbox{HITEMP-VENUS} \parencite{Pollack1993}. It is important to note that the hot lower atmosphere of Venus presents an additional computational challenge because the use of HITEMP or \mbox{HITEMP-VENUS} databases is generally required over HITRAN. This results in an increase of spectral lines by two orders of magnitude for CO$_2$ absorption modeling \parencite{Haus2010}. MARFA's line-by-line scheme shows capability of handling extensive line lists, like HITEMP, what we successfully employed for Venus studies.

\qquad Infrared continuum absorption parameters have only been established in specific transparency windows \parencite{Kappel2012, Kappel2016, Bezard2009, Bezard2011}, thus leaving the continuum function unavailable for broader spectral intervals in radiative transfer modeling. As a result, contributions from distant line wings must be accounted for. Various $\chi$-factors have been tested to describe the CO$_2$ sub-Lorentz behavior of spectral lines under Venusian conditions \parencite{Burch1969, fukabori1986chi-factor, bezard1990chi-factor, perrin1989chi-factor, Pollack1993, Tonkov1996, bezard2009chi-factor}, as well as different line cut conditions, typically ranging from 125 to 500 cm$^{-1}$ \parencite{Haus2010, Haus2013selfconsistent, Haus2015}. While using MARFA we systematically analyzed impact of these $\chi$-factors and line cut-offs on CO$_2$ absorption cross sections especially in Venus lower atmosphere. This explains benefits of using MARFA in sensitivity studies of Venus radiative transfer under varying spectroscopic settings. 

\begin{figure}
  \centering
    \includegraphics[scale=0.8]{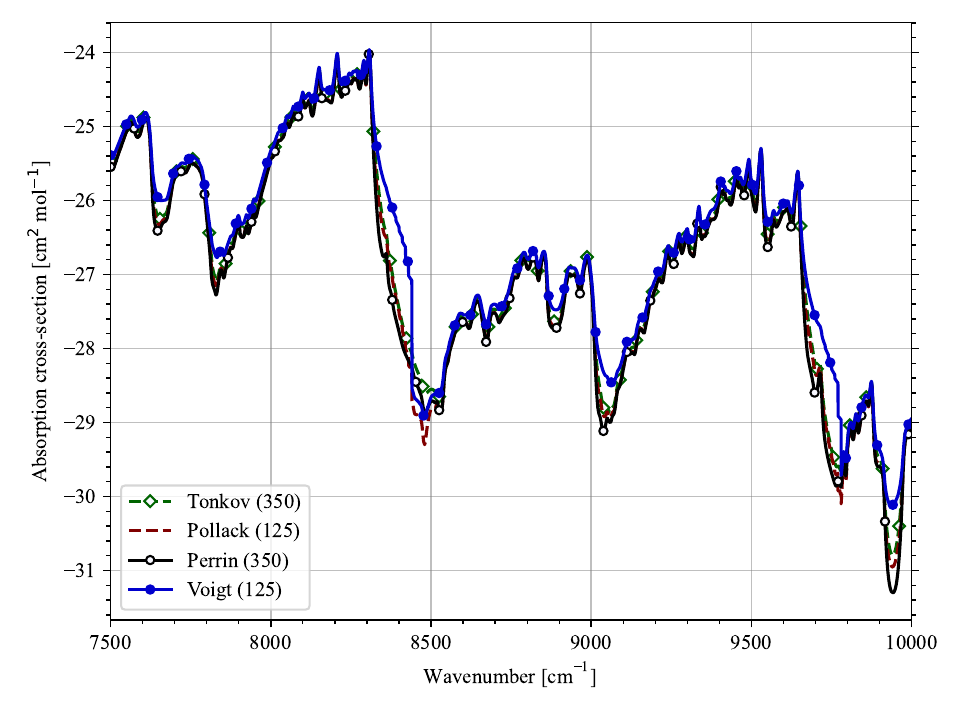}
  \caption{Monochromatic absorption cross-sections of CO$_2$ at 92.1 bar and 735.3 K as a function of wavenumber in dependence on sub-Lorentzian profile and line-cut condition. Continuum absorption is neglected.}
  \label{fig:chi-factors}
\end{figure}

\qquad On the fig.~\ref{fig:chi-factors} we present monochromatic absorption cross-sections of CO$_2$ at Venus surface calculated with MARFA for several line cut-off conditions and $\chi$-factors (\textcite{Tonkov1996, perrin1989chi-factor, Pollack1993}). The blue line represents default Voigt profile, meaning no line wing correction is applied. The fig.~\ref{fig:chi-factors} is validated against the fig.~10 in \textcite{Haus2010}, and shows good fits. Some discrepancies could be found in areas of low absorption most likely due to the choice of different spectral database and spectral resolution. These spectra are generated in high-resolution of $5\cdot10^{-4}$ cm$^{-1}$ and are based on HITRAN2020 line data \parencite{Gordon2022}. In the sensitivity study by \textcite{Haus2015} it has been confirmed that the choice of spectral line database (the most important), sub-Lorentz corrections and line cutoffs are essential for accurate Venus radiative balance modeling. Employing MARFA code allows to generalize this approach, enabling systematic exploration of databases/line shapes/cutoff impacts across diverse conditions.

\section{Implementation}
\label{sec:implementation}

\subsection{Code and project design}

\qquad In developing MARFA, we address the need for more flexible, data-driven, and efficient general-purpose line-by-line tools. However, specific attention is given to facilitate codes utilization: by both researchers (who just need to run the code) and by developers (who need also to contribute to the code). Scientific codes are known to be poorly adapted for community development, because of their complexity, lack of support, inconsistent coding standards and unrefactored legacy components. Such codes, even if being efficient in algorithmic terms, after distribution are considered often considered as black boxes, complicating the introduction of even small changes. In regard of targeting this issue, we highlight recently released codes for calculating atmospheric absorption by \textcite{korkin2025practical}. The authors of this study follow the so-called ``paper-and-code'' paradigm, when the paper extensively explains the code, and the code is openly available to complement the paper. Paper gives instructions not only how to run, but also how to change and develop the code. Our approach is closely aligned with this paradigm, although we moved more specific details to the well-structured documentation on the repository home page. Overall, MARFA might be understood as a kernel (or framework), which contains line-by-line scheme, basic line shapes, far wings correction functions and line intensity functions. This kernel is designed to be extensible and reusable under various atmospheric and spectroscopic conditions.

\qquad The MARFA project is thoroughly documented to provide quick-start instructions, detailed I/O formats and performance benchmarks \parencite{MARFA_repo}. The codebase employs intuitive variable names. Building and running the source code is straightforward using fpm (Fortran Package Manager) \parencite{fpm}. MARFA is designed through the Fortran modular structure. It means that each module within the source code is dedicated to specific physical or algorithmic operation, such as calculating line shapes, $\chi$-factors, grid calculations, etc. This division to modules improves clarity of the code and allows during development to focus on individual components without a risk affecting the entire program. During development and refactoring of the legacy code, we aimed to avoid using outdated features from older versions and followed the recent Fortran 2023 standard \parencite{metcalf2024modern, fortran2023}. Although the module \texttt{LineGridCalc.f90}, containing implementation of the multi-grid interpolation algorithm, still has legacy features, it does not compromise the overall clarity and maintainability of the code.

\qquad Following the surge in popularity of Python within the scientific community, several Python packages have been developed to calculate absorption spectra, either independently or as part of radiative transfer modeling schemes. Studies show (e.g. \textcite{schreier2018voigt}) that NumPy library \parencite{NumPy} provides comparable efficiency to Fortran or C/C++ languages. Notable examples include HAPI \parencite{HAPI}, py4Cats \parencite[]{schreier2019py4cats} (a Python reimplementation of GARLIC \parencite{GARLIC}), PyRTlib \parencite{PyRTlib}, among others. Nevertheless, Fortran was chosen as a main programming language of MARFA, primarily due to the existence of initial legacy code written in that language. Our approach offers two modes of interaction: through the well-documented Fortran source code and via the web platform \href{https://marfa.app}{marfa.app} which allows users to perform calculations directly by interacting with a lightweight version of the kernel. We may still later consider to convert MARFA or some of its modules (e.g. \texttt{LineGridCalc.f90} which contains implementation of the multi-grid algorithm) into a Python package even more facilitating development and distribution.

\subsection{Used data}
\qquad The tool is designed to minimize the amount of data stored within the project directory, with even fewer datasets embedded directly in the source code. Below we provide details about the datasets used in the project.
\subsubsection{Spectral Databases}
\label{subsec: databases}
\qquad The line-by-line approach implies accessing line data from spectral database files during runtime. For cross-section calculations (or absorption coefficients), only the essential (``core'') line parameters are necessary (see, e.g., \textcite{schreier2019py4cats}). To enhance efficiency, the MARFA executable expects these parameters to be stored in a binary file with a specified format. The source code offers built-in functionality to convert standard \texttt{.par} files into a required binary format. For more details, please refer to the \href{https://github.com/Razumovskyy/MARFA?tab=readme-ov-file\#spectral-databases}{project documentation} \parencite{MARFA_repo}.

\qquad Spectral databases for the first 12 molecules (based on HITRAN ID), precalculated to the needed format, are available in the \texttt{data/databases/} directory. These files are generated from \texttt{.par} files from HITRAN 2020. These files can be used to perform instant calculations or testing. They cover the range from 10 cm$^{-1}$. By knowing the required file format (see appendix \ref{apx: utility for processing}), users can generate and store their own spectral databases for repeating calculations. By providing database parameter when running MARFA executable, it is easy to switch between different databases and with that make a comparative analysis. In the next releases we plan to introduce support of the near coming HITRAN 2024 database.

\subsubsection{Total Internal Partition Sums (TIPS)}
\qquad Total internal partition sums (TIPS) are needed to determine temperature-dependent spectral line intensities. In MARFA, TIPS data are set in an external file and are implemented based on the study by \textcite{Gamache2016}. We have introduced only partial data covering the first 74 isotopologues for the first 12 molecules (based on HITRAN ID and isotopologue local ID), over a temperature range of 20–1000 K. The TIPS values for a given isotopologue as a function of temperature can be accessed via the function \texttt{TIPSofT} in the \texttt{Spectroscopy.f90} module. Future releases will incorporate more recent data from \textcite{Gamache2021}, for more broad range of molecules and with considerations for reorganizing the input structure using Fortran subroutine or Python module.

\subsubsection{Molecular Weights}
\qquad Molecular weights are used for calculating Doppler half-widths and are stored in the \texttt{MolarMasses.f90} module for the variety of isotopologues. 

\subsection{Quick start instructions}

\qquad The process involves cloning the repository, building the project using the \texttt{fpm build} command, and installing the required Python packages. Predefined atmospheric profiles, such as \texttt{VenusCO2.dat} — which models the CO$_2$ distribution in Venus’s nightside atmosphere — are available for use. Currently, these profiles are limited to Venus constituents; alternatively, users may create custom atmospheric profiles following the specified format detailed in the project documentation \parencite{MARFA_repo}. 

\qquad Absorption coefficients are computed by executing the \texttt{fpm run} command with command line arguments that include the molecule title, spectral interval, line cut-off condition, atmospheric profile file, and other parameters. Post-processing is performed using an interactive provided Python script. Upon using it, a user is able to convert binary look-up table into human-readable format and to generate plot with dynamically adjusted title.

\subsection{Scope of application and limitations}

\qquad The codes are particularly well-suited for modeling absorption features in environments where both spectral and atmospheric parameters are uncertain. Our initial focus was made on Venus atmosphere, which explains current absence of predefined atmospheric profiles for other planets. The major scenario targeted by MARFA's eleven-grid interpolation technique involves handling large cut-offs of spectral lines when continuum absorption function is missing. For that reason, currently MARFA does not support the incorporation of continuum absorption.

\qquad Currently preprocessed spectral databases only for the first 12 molecules (based on HITRAN ID) are available out of the box. This is a temporary limitation and it is due to the large size of these databases, making it impractical to store them in the repository. For details on how to generate spectra for additional molecules, refer to subsection \ref{subsec: databases}. Only cross-sections or volume absorption coefficients can be calculated when running current version of MARFA and it does not include functionality for modeling instrumental effects. Another limitation is the absence of support of the microwave region. It is planned to be added later, by making specific adjustments, such as Clough correction \parencite{clough1980correctionVanVleck} to the van Vleck-Weisskopf (or van Vleck-Huber) theory \parencite{VanVleck1977} and handling ``negative wavenumbers''.

\subsection{Web application marfa.app}
\label{subsec:web-interface}

\begin{figure}[!htbp]
    \centering
    \includegraphics[width=0.5\linewidth]{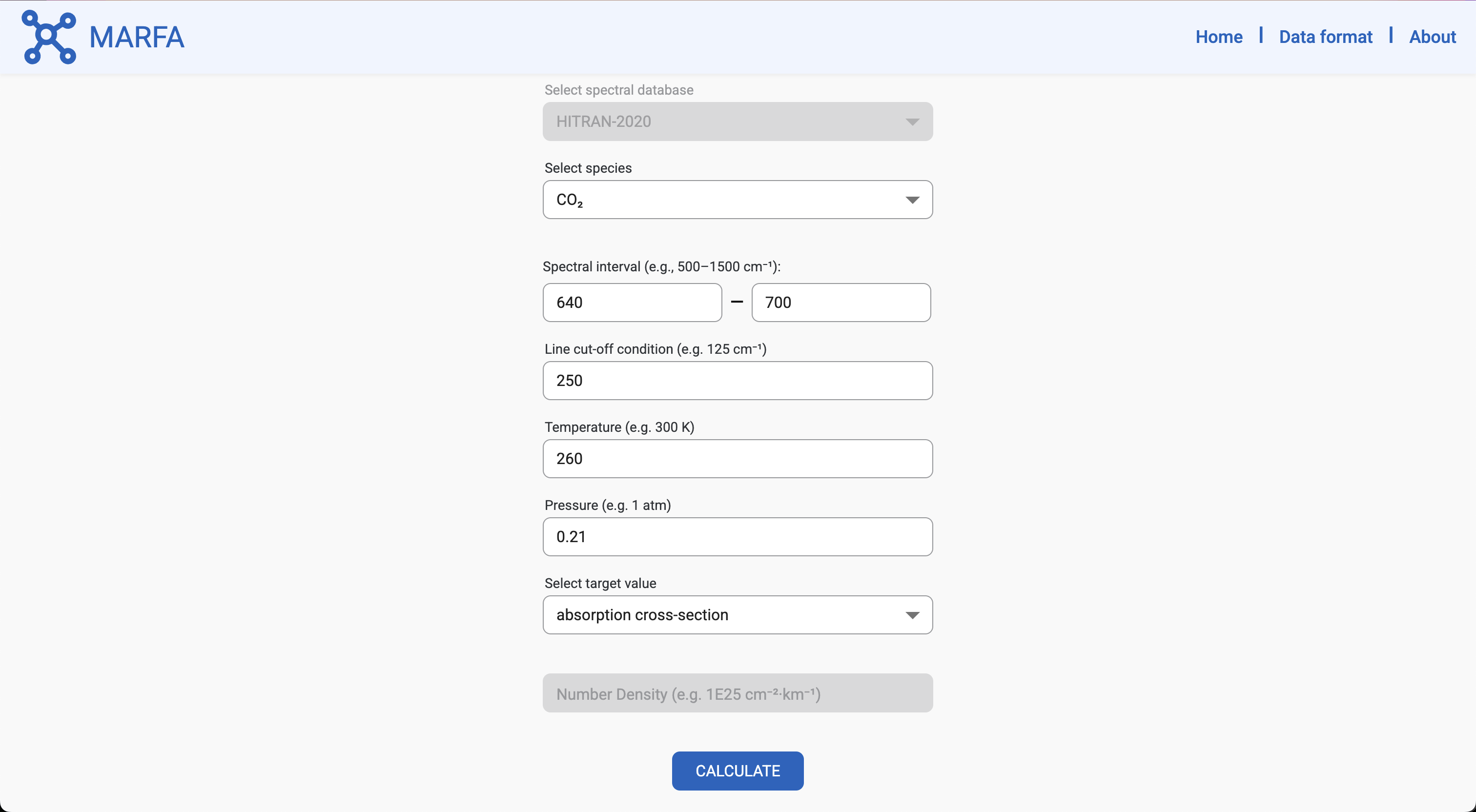}
    \caption{Interface for setting parameters}
    \label{fig2:web-interface1}
\end{figure}

\begin{figure}[!htbp]
    \centering
    \includegraphics[width=0.45\linewidth]{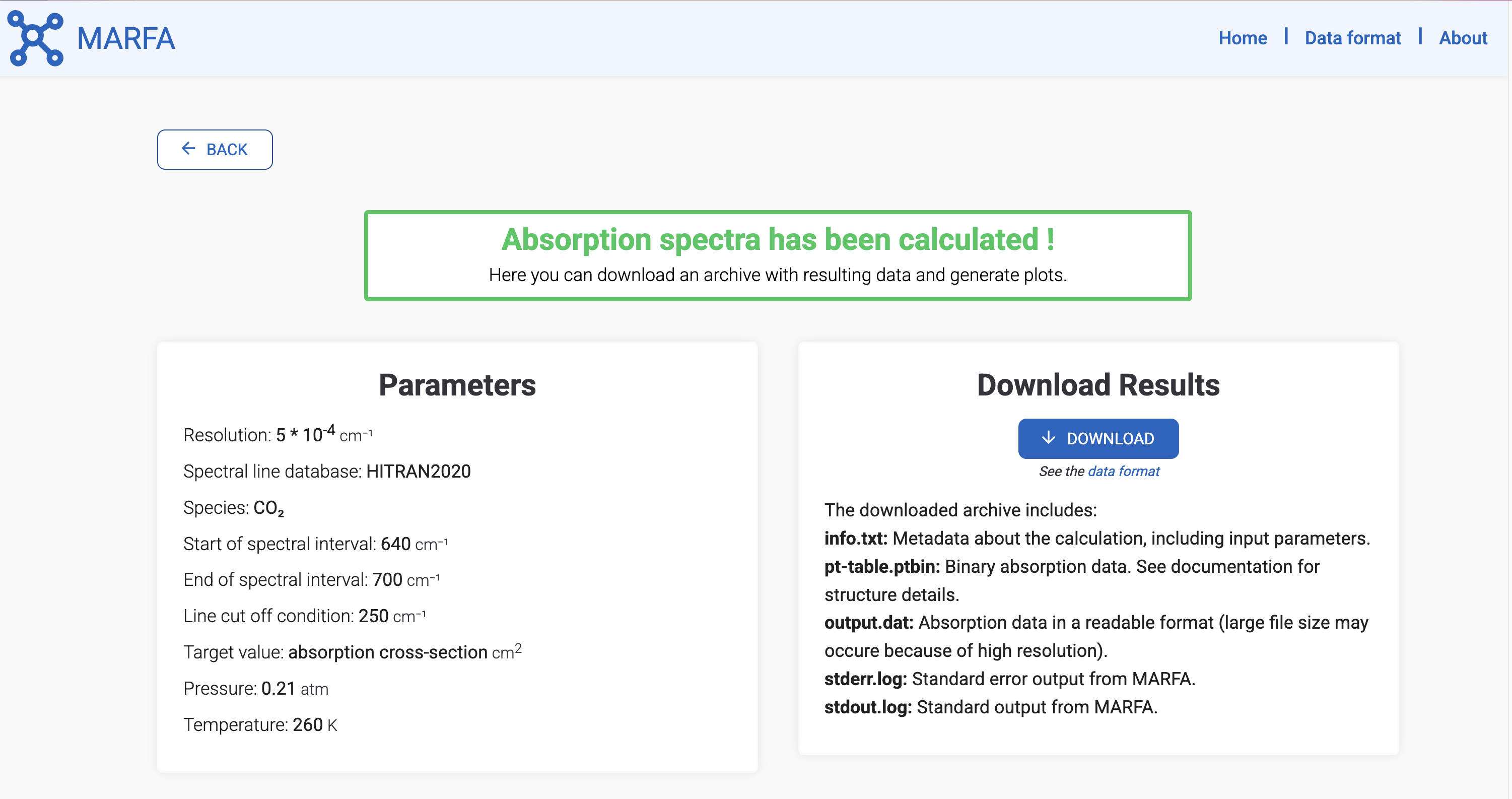}
    \caption{Interface for download options}
    \label{fig3:web-interface2}
\end{figure}

\begin{figure}[!htbp]
    \centering
    \includegraphics[width=0.45\linewidth]{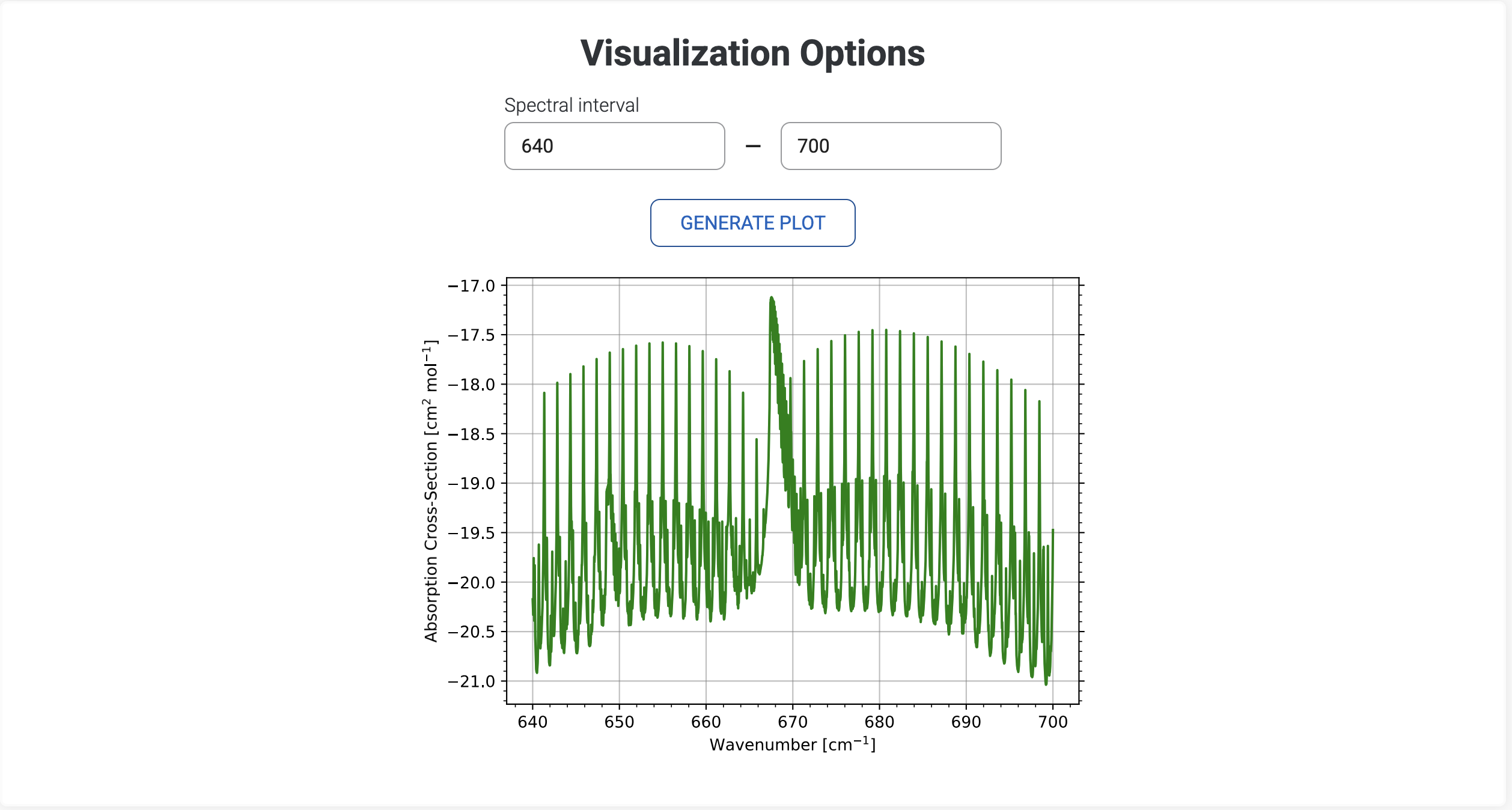}
    \caption{Interface for plot generation}
    \label{fig4:web-interface3}
\end{figure}

\qquad A \href{https://marfa.app}{web interface} accessible at \href{https://marfa.app}{marfa.app} has been developed enabling an immediate interaction with MARFA atmospheric absorption calculator. The example images ~\ref{fig2:web-interface1}, ~\ref{fig3:web-interface2} and ~\ref{fig4:web-interface3} illustrate essential parts of the application. We consider \href{https://marfa.app}{marfa.app} as a good sketch of an atmospheric spectroscopy calculation lab/platform, build with well-established, modern, scalable and robust web stack. It may be an example of how to improve, transform and scale in near future existing more rich-functional solutions, like \href{https://hitran.iao.ru/}{hitran.iao.ru} \parencite[]{HITRAN_on_the_WEB}. Moreover \href{https://marfa.app}{marfa.app} is lightweight and has an easy interface and could be used with ease in educational purposes. Web architecture of the marfa application allows to deploy it on any other hosting in without any issues. The source code is public and hosted on the dedicated repository \parencite{MARFA_webapp}, and like the MARFA source code it can be freely contributed and distributed under provided licenses. 

\qquad The client side is built using the Next.js framework \parencite{nextjs}, which supports quick server-side rendering. On the server side, Django \parencite{django} and Django Rest Framework \parencite{djangorestframework} backend frameworks are used, which are well-known and are highly scalable. They provide robust built-in database support mechanisms and functionality for building Web APIs (REST API is our primary choice). For the HTTP web server, Nginx \parencite{nginx} is used and PostgreSQL \parencite{postgresql} is used as a primary database for storing information about users requests.

\qquad Integration between the backend and MARFA calculation kernel is achieved by leveraging Django's ViewSet to invoke external processes via Python's subprocess module \parencite{python_subprocess}. This was chosen instead of \texttt{f2py} module from the Numpy \parencite{NumPy}, because there was a need to separate Fortran and Python layers of the application.  Upon completion, the Fortran code generates the resulting PT-tables files, which are stored on the server. After processing of these files is complete, they are provided to a user with download links. Frontend and backend parts of the application are hosted within docker \parencite{Docker} containers, and a CI/CD pipeline based on GitHub Actions \parencite{GitHub_Actions} is incorporated for automatic deployments.

\FloatBarrier

\section{Conclusion}
\label{sec:conclusion}

\qquad MARFA (Molecular atmospheric Absorption with Rapid and Flexible Analysis) addresses critical challenges in calculation of molecular absorption in planetary atmospheres under sparse spectroscopic and atmospheric data conditions. By implementing an eleven-grid interpolation technique \parencite{fomin1995effective}, MARFA can efficiently process large line cut-offs (500 $\text{cm}^{-1}$ and beyond) and maintain high spectral resolution ($\approx5\times10^{-4} \text{cm}^{-1}$). This approach significantly reduces computational costs and makes it suitable for planetary conditions like Venus, where continuum functions remain poorly characterized on large spectral intervals.

\qquad The MARFA tool was tested using Venus as a benchmark, addressing challenges such as high-temperature CO$_2$ absorption, uncertain atmospheric profiles, uncertain line data and the need for extensive line-wing corrections. MARFA's features govern the ability to rapidly recalculate absorption spectra under varying parameters, which can make it valuable in diverse scenarios from Venusian conditions to exoplanetary analogs.

\qquad Provided not only as source code but also with a companion web interface (\href{https://marfa.app}{marfa.app}), MARFA lowers the barrier to entry for educational and exploratory use. Because of its code design, effectiveness, flexibility and open-source commitment, MARFA project (source code + web application) can be a valuable resource for:
\begin{enumerate}
    \item \textit{Atmospheric spectroscopy research}: fast and accurate calculation of molecular spectra for radiative transfer models, general circulation models and in other frameworks; providing sensitivity analyses.
    \item \textit{Educational outreach}: teaching molecular spectroscopy and modern Fortran programming.
    \item \textit{Community-driven development}: transparent issue tracking and ease of installation and local development invite contributions from both scientific and IT-communities.
\end{enumerate}

\subsection*{Future directions}

\qquad Future improvements of MARFA are planned to be made by the authors of this study and by merge requests approvals of the independent contributions from the community. Planned improvements are tracked on the \href{https://github.com/Razumovskyy/MARFA/issues}{issues page}, aiming to make it clear for contributors what is the current state of the project and what are the directions of improvements. Key focus areas are:
\begin{enumerate}
    \item \textit{Community-driven development}: Actively seeking for a feedback from researchers to test MARFA in novel scenarios; we hope that community input will guide us to prioritize new features.
    \item \textit{Performance Optimizations}:   
    \begin{itemize}
        \item Parallelizing spectral calculations across atmospheric levels using OpenMP \parencite{openmp} (inspired by  \textcite{GARLIC})
        \item Reducing I/O bottlenecks in line-by-line processing through optimizing nested \texttt{DO} loops.  
        \item Introducing dynamic or user-defined spectral resolution.  
    \end{itemize}  
    \item \textit{Data expansion}:
    \begin{itemize}
        \item Integration and support of the HITRAN 2024 database and custom spectroscopic datasets (e.g., high-temperature CO$_2$ profiles). This may also include initializing a data storage for spectral databases preprocessed for MARFA calculations.
        \item Enhancing TIPS datasets support.  
        \item Expanding the set of predefined atmospheric profiles (e.g. for Mars, Titan). 
    \end{itemize}  
    \item \textit{Codebase Modernization}:  
    \begin{itemize}
        \item Refactoring legacy module \texttt{LineGridCalc.f90} to adhere to the modern Fortran standards. We also consider converting this module, containing implementation of the interpolation method \parencite{fomin1995effective} to the Python package.
        \item Resolving bugs identified through testing.  
    \end{itemize}  
    \item \textit{New Features}:  
        \begin{itemize}
            \item Adding apparatus function convolution for instrumental response simulations.  
            \item Incorporating support of continuum absorption functions .
            \item Extending MARFA support to microwave regimes.
        \end{itemize}  
    \item \textit{Web platform}: Improvements and scaling of the web platform relied on the needs of researchers and educational collectives.
\end{enumerate}

\newpage

\section*{Appendix A: Structure of the output PT-table file}

\qquad The output PT-table files are generated and stored within the \texttt{output/ptTables} directory. Each PT-table file is a binary file and corresponds to a specific atmospheric level on which absorption was calculated. File name convention features only the level number, e.g. \texttt{1\_\_.ptbin} or \texttt{65\_.ptbin}. All the output files have the extension \texttt{.ptbin}.

\qquad The structure of the output file is organized into records that contain high-resolution absorption data (direct access file). Each record follows the structure outlined below:

\subsection*{Absorption data and resolution}
 \qquad Each record stores an array, which contains either the cross-section or absorption coefficient data for a specific wavenumber range.
 Each record corresponds to a 10 cm$^{-1}$-length interval, controlled by the \texttt{deltaWV} parameter in the source code. The finest grid in the interpolation technique \parencite{fomin1995effective} contains \texttt{NT} = 20481 nodes, thus defining the resolution of the output data: 

    \[
    \delta\nu = \frac{\texttt{deltaWV}}{\texttt{NT-1}} = \frac{10}{20480} \approx 5 \times 10^{-4} \, \text{cm}^{-1}
    \]

\subsection*{Record Indexing}
\qquad The relationship between a given wavenumber $\nu$ and the record number is defined as:
    \[
    \texttt{recNum} = \texttt{int}\left(\frac{\nu}{10}\right)
    \]
\qquad For example, if absorption data for 7564 cm$^{-1}$ is required, the corresponding record would be the 756th entry. For more details we address readers to the documentation \parencite{MARFA_repo}, where additionally a code snippet is provided to extract data from the PT-table.

\section*{Appendix B: Utility for processing .par files}
\label{apx: utility for processing}

\qquad MARFA requires spectral database files to be in a specific format. These files should be placed in the \texttt{data/databases} directory and are typically created based on \texttt{.par} files from sources such as the HITRAN database. These \texttt{.par} files need to be preprocessed into required format before running the main executable. For users convenience, the utility provided in the codebase allows users to perform this preprocessing with simple command, and placing generated files in a required directory. For more details, please refer to the \href{https://github.com/Razumovskyy/MARFA?tab=readme-ov-file#spectral-databases}{spectra databases} section in the documentation \parencite{MARFA_repo}.

\subsection*{Database file structure}

\qquad Each spectral database file is a binary file which is organized as a collection of records (like the output PT-table file). Each record corresponds to one spectral line and contains its ``core'' parameters. The parameters include (for more details for each item, see \textcite{hitran_definitions_units}): 
\begin{enumerate}
    \item the wavenumber of the transition ($\nu_{ij}$) in double precision
    \item the spectral line intensity ($S_{ij}$) at 296 K
    \item the air-broadened half width at half maximum (HWHM) at 296 K ($\gamma_{\text{air}}$)
    \item the self-broadened half width at half maximum (HWHM) at 296 K ($\gamma_{\text{self}}$)
    \item the lower-state energy of the transition ($E^{\prime\prime}$)
    \item the coefficient of the temperature dependence of the air-broadened half width ($n_{\text{air}}$)
    \item \texttt{jointMolIso} - custom parameter, which contains information on both molecule id (Mol) and isotopologue id (Iso). For more details, refer to the documentation.
    \item the pressure shift at 296 K and 1 atm of the line position with respect to the vacuum transition wavenumber ($\delta_{\text{air}}$) 
\end{enumerate}
Since there are 8 parameters and for storing transition wavenumber one 8 bytes are needed, the total record length equals to $7\times4 + 8 = 36$ bytes.

\section*{Appendix C: Implementation of Voigt function}
\label{appx: Voigt}
\qquad The approach we follow to compute the Voigt function is based on Humlicek \parencite{Humlicek1982} and Kutnz's \parencite{Kuntz1997} works. Original Kuntz's study optimizes Humlicek's method in two ways. Firstly, it rewrites complex rational approximations as real arithmetic operations, removing imaginary part computations unnecessary in applications, thus reducing floating-point operations. Secondly, it introduces recursive algorithm that leverages sorted x-grids to process subintervals collectively, reducing per-point region checks.

To summarize, our implementation incorporates the Kuntz's approach but with several modifications:
\begin{enumerate}
    \item For $x > 15$, we substitute Voigt function with Lorentzian wing (optionally corrected with $\chi$-factor) without fall in accuracy.
    \item Kuntz's rational approximations are adopted for regions 1-3 (see fig.1 and Appendix in \parencite{Kuntz1997}).
    \item Region 4 boundaries in $(x,y)$-plane are slightly changed and are set to: $\{y<0.02, 1<x<5.5\}$ for an optimal speed-accuracy trade-off. Due to the Voigt function converging to Lorentzian for large $x$, we found it more convenient to use direct analytical asymptotic expressions instead of cumbersome polynomials  (see pp. 823-824 in \parencite{Kuntz1997}), maintaining relative errors below 1\%.
    \item Recursive subdivision algorithm is omitted in favor of  direct \texttt{if-else} blocks for region detection per (x,y) pair. We found it preferable to use simple and reliable if-else blocks instead of recursion due to the high vectorization capabilities of modern CPUs.
\end{enumerate}

Below there is a demonstration of one of the asymptotics we use in region 4.
\begin{equation}
\label{eq: K-function}
    K(x,y) = \frac{y}{\pi}\int_{-\infty}^{+\infty}e^{-t^2}\phi(t),\qquad
    \text{where}\qquad \phi(t) = \frac{1}{(x-t)^2 + y^2}
\end{equation}
is a function that varies significantly on a characteristic scale of $y$ and $y \ll 1$.

To get analytical approximation, Taylor expansion to the second order term is applied:
\begin{equation}
\label{eq: K-taylor}
    \frac{\pi}{y}K(x,y) = \phi(0)\int_{- \infty}^{+ \infty} e^{-t^2} dt + \frac{\partial \phi}{\partial t}\int_{- \infty}^{+ \infty} e^{-t^2} t dt  + 
     \frac{\partial^2 \phi}{\partial t^2}\int_{- \infty}^{+ \infty} e^{-t^2} t^2 dt
\end{equation}
In the expression \ref{eq: K-taylor}, first integral is $\sqrt{\pi}$, second integral is zero, while third one equals to \mbox{$\frac{1}{2}\Gamma(\frac{3}{2})=\frac{\sqrt{\pi}}{4}$}. After passing values and derivatives of $\phi(t)$ to the \ref{eq: K-function} and \ref{eq:voigt_line_shape}, we get the asymptotic expression for the Voigt function:
\begin{equation}
    f_{\text{V}} \approx f_{\text{L}}\cdot\left(1 + \frac{3}{2x^2}\right).
\end{equation}


\newpage

\printbibliography

\end{document}